\documentclass[conference]{IEEEtran}
\IEEEoverridecommandlockouts
\usepackage{cite}
\usepackage{amsmath,amssymb,color,graphicx,caption,subcaption,comment,tikz,url,latexsym} 
\usepackage{algorithmic}
\usepackage{graphicx}
\usepackage{textcomp}
\usepackage{pgfplots}
\usepackage{xcolor}
\usepackage{url}
\usepackage{multirow}
\usepackage{mathtools}
\usepgfplotslibrary{units}
\usetikzlibrary{spy,backgrounds}
\usepackage{pgfplotstable}
\usepackage[left=0.6in,right=0.6in,top=0.7in,bottom=0.97in]{geometry}

\def\BibTeX{{\rm B\kern-.05em{\sc i\kern-.025em b}\kern-.08em
		T\kern-.1667em\lower.7ex\hbox{E}\kern-.125emX}}

\newtheorem{thm}{Theorem}
\newtheorem{example}{Example}
\newtheorem{definition}{Definition}
\newtheorem{lemma}{Lemma}

\newtheorem{corollary}{Corollary}

\usepackage{bbm}

\newcommand{\M}{\mathsf{M}}
\newcommand{\m}{\mathsf{m}}
\newcommand{\len}{\textrm{len}}

\newcommand{\E}{\mathbb{E}}

\begin{document}
	\interdisplaylinepenalty=0
	\title{Benefits of Rate-Sharing for Distributed Hypothesis Testing\\
	}
	\author{\IEEEauthorblockN{Mustapha Hamad}
		\IEEEauthorblockA{\textit{LTCI, Telecom Paris, IP Paris} \\
			91120 Palaiseau, France\\
			mustapha.hamad@telecom-paris.fr}

		\and
		\IEEEauthorblockN{Mireille Sarkiss}
		\IEEEauthorblockA{\textit{SAMOVAR, Telecom SudParis, IP Paris} \\
			91011 Evry, France\\
			mireille.sarkiss@telecom-sudparis.eu}
				\and
				\IEEEauthorblockN{Mich\`ele Wigger}
				\IEEEauthorblockA{\textit{LTCI, Telecom Paris, IP Paris} \\
					91120 Palaiseau, France\\
					michele.wigger@telecom-paris.fr}
	}
	\allowdisplaybreaks[4]
	\sloppy
	\maketitle

	\begin{abstract}
We study  distributed binary hypothesis testing with a single  sensor and two remote decision centers that are also equipped with local sensors. The communication between the  sensor and the two decision centers takes place over three links: a shared link to both centers and  an individual link to each of  the two  centers. All communication links are subject to \emph{expected} rate constraints. This paper characterizes the optimal exponents region of the type-II error for given type-I error thresholds at the two decision centers and further simplifies the expressions in the special case of having only the single shared link. The exponents region illustrates a gain under expected rate constraints compared to equivalent maximum rate constraints. Moreover, it exhibits a tradeoff between the exponents achieved at the two centers.
	\end{abstract}
	\begin{IEEEkeywords}
	Broadcast channel, distributed hypothesis testing, error exponents, expected rate constraints, IoT, decision centers.
	\end{IEEEkeywords}	
	\section{Introduction}

We address a distributed hypothesis testing problem where different decision centers have to decide on the same  hypothesis based on their local sensing and the messages they receive from remote sensors over rate-limited communication links. 
Motivated by systems that share bandwidth among several applications with variable instantaneous bandwidth for each application, 
we consider \emph{expected-rate constraints} that limit only the  expected bandwidth for each application.

In our work, we focus on distributed binary hypothesis testing against independence. The decision centers have to decide between a \emph{i) null hypothesis} (normal situation) indicating that  the centers' and the sensors' observations are correlated, and an \emph{ii) alternative hypothesis} (alert situation) where the observations are independent, for example because one of the systems fails. Two types of errors can be distinguished: the \emph{type-I error} indicates a wrong decision under the null hypothesis and the \emph{type-II error} occurs if a wrong decision is made under the  alternative hypothesis. Since the alternative hypothesis corresponds to a more critical situation, we aim at maximizing the exponential decay of the type-II error probability, called \emph{error exponent}, subject to   a type-I error that stays below a given threshold.
Such a  setup has been studied in many previous works focusing mostly on \emph{maximum-rate constraints}\cite{Ahlswede,Han,Amari,Wagner,Michele2,PierreMichele,zhao2018distributed,Michele_BC,Kim,Watanabe_DHT,Kochman-MAC,Kochman,Deniz_DHT_Privacy,Deniz_DHT,Vincent_Tan_DHT_Privacy,Tyagi_DHT}. \emph{Expected-rate constraints} were introduced in \cite{JSAIT}, where the maximum error exponent  for single-sensor single-decision center setup was characterized in the special case of testing-against independence. Extensions of this work were first proposed for a multi-sensor scenario in \cite{ITW20}, for a multi-hop scenario with multiple decision centers in \cite{GLOBECOM21,ITW21}, and most recently from a signal detection perspective in \cite{Telatar_DHT_Signal_Detection}.

In this paper, we consider a single-sensor  two-decision center scenario where the decision centers also have sensing capabilities.  The communication takes place over three noise-free links: a common link to both decision centers and one private link to each decision center. 
For this one-to-many broadcast setup, we characterize the  optimal exponents region under expected-rate constraints and we show that it improves over the exponents region under maximum-rate constraints, which we also establish in this paper. 
The  optimal exponents region under expected rate constraints illustrates two tradeoffs. The first tradeoff results from the shared link that has to serve both decision centers at the same time; this tradeoff is also present under maximum-rate constraints.  The second tradeoff is particular to the setup with expected-rate constraints and stems from the rate-sharing between  three different variants   of the optimal coding scheme under maximum-rate constraints in  \cite{Michele_BC}, depending on the observations at the sensor. We show that two variants suffice when communication is only over a single shared link,
 leading to significant reduction in the complexity of the optimal coding scheme.

\textit{Notation:}
We follow the notation in \cite{ElGamal},\cite{JSAIT}. In particular, we use sans serif font for bit-strings: e.g., $\m$ for a deterministic and $\M$ for a random bit-string, and we denote the length of $\m$ by $\mathrm{len}(\m)$. 
 In addition, $\mathcal{T}_{\mu}^{(n)}(P)$ denotes the strongly $\mu$-typical set  with respect to $P$ as defined in \cite[Definition 2.8]{Csiszarbook}.

	\section{System Model}
	
\begin{figure}[ht]
	\centerline{\includegraphics[ scale=0.15]{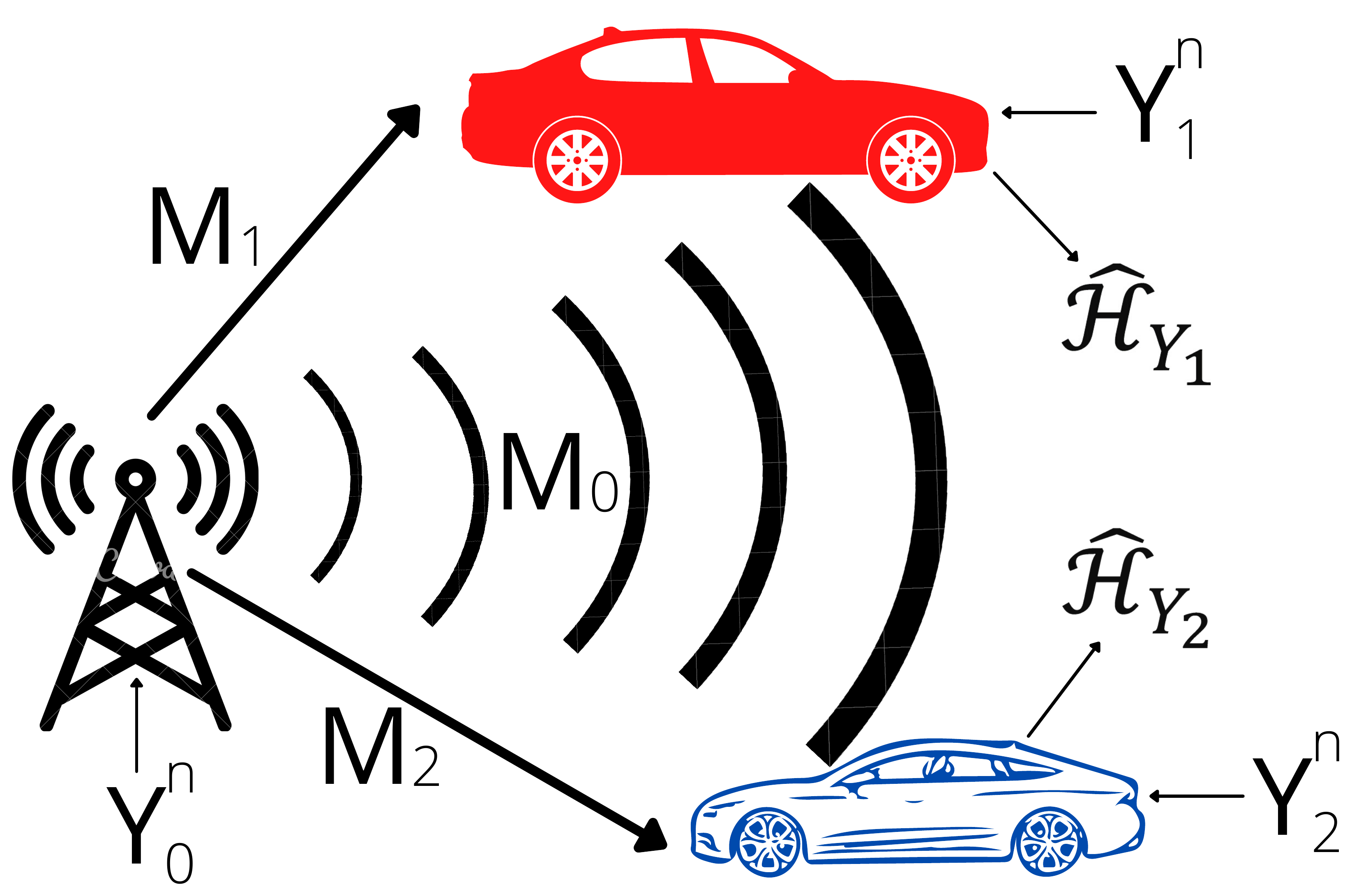}}
	\vspace{2mm}
	
	\caption{Distributed hypothesis testing with a single  sensor and two remote decision centers with integrated sensors.}
	\label{fig:BC}\vspace{-0.15cm}
\end{figure}

	Consider the distributed hypothesis testing problem in Figure~\ref{fig:BC} in the special case of testing against independence, i.e., depending on the binary hypothesis $\mathcal{H}\in\{0,1\}$, the tuple $(Y_0^n,Y_1^n,Y_2^n)$ is distributed as:
	\begin{subequations}\label{eq:dist}
		\begin{IEEEeqnarray}{rCl}
			& &\textnormal{under } \mathcal{H} = 0: (Y_0^n,Y_1^n,Y_2^n) \sim \textnormal{i.i.d.} \, P_{Y_0}\cdot P_{Y_1Y_2|Y_0} ; \label{eq:H0_dist}\IEEEeqnarraynumspace\\
			& &\textnormal{under } \mathcal{H} = 1: (Y_0^n,Y_1^n,Y_2^n) \sim \textnormal{i.i.d.} \, P_{Y_0}\cdot P_{Y_1Y_2}
		\end{IEEEeqnarray} 
	\end{subequations}
	for given probability mass functions (pmfs) $P_{Y_0}$ and $P_{Y_1Y_2|Y_0}$ and where $P_{Y_1Y_2}$ denotes the marginal of the joint pmf $P_{Y_0Y_1Y_2}:=P_{Y_0}P_{Y_1Y_2|Y_0}$. 

	The system consists of a transmitter T$_{Y_0}$, and two receivers R$_{Y_1}$, R$_{Y_2}$. Transmitter T$_{Y_0}$ observes the source sequence $Y_0^n$ and computes three bit-string messages $(\M_0,\M_1,\M_2) = \phi^{(n)}(Y_0^n)$, where the encoding function is of the form $\phi^{(n)} : \mathcal{Y}_0^n \to \{0,1\}^{\star} \times \{0,1\}^{\star} \times \{0,1\}^{\star}$. Message $\M_0$ is sent to both receivers R$_{Y_1}$, R$_{Y_2}$,  while  message $\M_1$  only to  receiver R$_{Y_1}$ and message $\M_2$ only to receiver R$_{Y_2}$. The messages have to satisfy the \emph{expected} rate constraints
	\begin{equation}\label{eq:Ratei}
		\mathbb{E}\left[\mathrm{len}\left(\M_i\right)\right]\leq nR_i, \qquad i \in \{0,1,2\}.
	\end{equation} 
	Receiver R$_{Y_i}, i \in \{1,2\},$ observes the source sequence $Y_i^n$ and with  messages $\M_0, \M_i$ received from T$_{Y_0}$, it produces a guess  $\hat{\mathcal{H}}_{Y_i}$ of the hypothesis ${\mathcal{H}}$ using a decision function $g_i^{(n)} : \mathcal{Y}_i^n \times \{0,1\}^{\star} \times \{0,1\}^{\star} \to \{0,1\}$:
	\begin{equation}
		\hat{\mathcal{H}}_{Y_i} = g_i^{(n)}\left(Y_i^n,\M_0,\M_i\right) \;   \in\{0,1\}, \qquad i \in \{1,2\}.
	\end{equation}
	
	The  goal is to design encoding and decision functions such that their type-I error probabilities 
	\begin{equation}
		\alpha_{i,n} \triangleq \Pr[\hat{\mathcal{H}}_{Y_i} = 1|\mathcal{H}=0],  \qquad i \in \{1,2\},
	\end{equation}
stay below given thresholds $\epsilon_i > 0,  i \in \{1,2\}$, 
and the type-II error probabilities
		\begin{equation}
		\beta_{i,n} \triangleq \Pr[\hat{\mathcal{H}}_{Y_i} = 0|\mathcal{H}=1]
	\end{equation}
	decay to 0 with largest possible exponential decay.

	\begin{definition} Fix maximum type-I error probabilities $\epsilon_1,\epsilon_2 \in [0,1]$ and rates $R_1,R_2 \geq 0$. The exponent pair $(\theta_1,\theta_2)$ is called \emph{$(\epsilon_1,\epsilon_2)$-achievable} if there exists a sequence of encoding and decision functions $\{\phi^{(n)},g_1^{(n)},g_2^{(n)}\}_{n\geq 1}$ satisfying:
		\begin{subequations}
			\label{eq:RPconstraints}
		\begin{IEEEeqnarray}{rCl}
			\mathbb{E}[\text{len}(\M_i)] &\leq& nR_i, \quad i \in \{0,1,2\} \label{eq:LEN}\\
			\varlimsup_{n \to \infty}\alpha_{i,n} & \leq& \epsilon_i, \qquad i \in \{1,2\} \label{type1constraint1}\\ 
			\label{thetaconstraint}
			\varliminf_{n \to \infty}  {1 \over n} \log{1 \over \beta_{i,n}} &\geq& \theta_i, \qquad i \in \{1,2\}.
		\end{IEEEeqnarray}
				\end{subequations}
	\end{definition}
	\smallskip
	
	\begin{definition}
		The closure of the set of all $(\epsilon_1,\epsilon_2)$-achievable exponent pairs $(\theta_{1},\theta_{2})$ is called the \emph{$(\epsilon_1,\epsilon_2)$-exponents region} and is denoted  $\mathcal{E}^*(R_0,R_1,R_2,\epsilon_1,\epsilon_2)$. 
		
	\end{definition}
	
	\section{Main Results}
	Our main results are a complete characterization of the exponents region $\mathcal{E}^*(R_0,R_1,R_2,\epsilon_1,\epsilon_2)$ under the \emph{expected}-rate constraints in \eqref{eq:Ratei} as well as a strong converse under analogous \emph{maximum}-rate constraints. 
	A simplified expression is provided for $\mathcal{E}^*(R_0,0,0,\epsilon_1,\epsilon_2)$.
	
	\subsection{Individual and Common Communication Links}
		
		\begin{thm}\label{thm1}
			The $(\epsilon_1,\epsilon_2)$-exponents region $\mathcal{E}^*(R_0,R_1,R_2,\epsilon_1,\epsilon_2)$ \emph{is the set of} all ($\theta_{1},\theta_{2}$) pairs  satisfying 
			\begin{subequations}\label{eq:E3}
				\begin{equation}\label{eq:thetaE3}
					\theta_{i} \leq \min\left\{ I\big(U_0^{0}U_i^{0};Y_i\big), I\big(U_0^{i}U_i^{i};Y_i\big)\right\}, \qquad i \in \{1,2\}
				\end{equation}
				for some non-negative numbers $\sigma_0,\sigma_{1},\sigma_2$  with sum $\leq 1$ and conditional pmfs $P_{U_0^0|Y_0}$, $P_{U_0^1|Y_0}$, $P_{U_0^2|Y_0}$, $P_{U_1^0|U_0^0Y_0}$, $P_{U_1^1|U_0^1Y_0}$, $P_{U_2^0|U_0^0Y_0}$, $P_{U_2^2|U_0^2Y_0}$ satisfying
				\begin{IEEEeqnarray}{rCl}
					R_0& \geq & \sigma_{0}I(U_0^{0};Y_0) + \sigma_{1} I(U_0^{1};Y_0) + \sigma_{2}I(U_0^{2};Y_0), \label{eq:R0une} \IEEEeqnarraynumspace\\
					R_i& \geq & \sigma_{0}I(U_i^{0};Y_0|U_0^{0}) + \sigma_{i} I(U_i^{i};Y_0|U_0^{i}), \quad i\in\{1,2\}, \IEEEeqnarraynumspace \label{eq:R2une} \vspace{-0.4cm}
			\end{IEEEeqnarray}			
		\begin{IEEEeqnarray}{rCl}\label{eq:sigmas}
	         \text{and} \quad  \sigma_0+ \sigma_i & \geq& 1- \epsilon_i, \quad i\in\{1,2\},\\
		  \sigma_0 &  \geq & 1- \epsilon_1-\epsilon_2,\label{eq:sigmas2}
		\end{IEEEeqnarray}	
			\end{subequations}
		and where the mutual information quantities are calculated according to the joint pmfs 
		\begin{IEEEeqnarray}{rCl} 
		P_{Y_0Y_1Y_2U_0^0U_1^0U_2^0} &\triangleq & P_{Y_0Y_1Y_2}
		P_{U_0^0|Y_0}P_{U_1^0U_2^0|U_0^0Y_0}\\
		P_{Y_0Y_1Y_2U_0^iU_i^i}  &\triangleq & P_{Y_0Y_1Y_2}P_{U_0^i|Y_0}P_{U_i^i|U_0^iY_0}, \quad i\in\{1,2\}. 
				\end{IEEEeqnarray}	
		\end{thm}
		\begin{IEEEproof}  The achievability is proved in Appendix~\ref{app:Ach.}. The converse is proved in Section~\ref{general_converse}. 
		\end{IEEEproof} 
		\smallskip
	Theorem~\ref{thm1} shows a tradeoff between the two achievable exponents $\theta_1$ and $\theta_2$. 	(Figure~\ref{fig:BC_Trenary_VL_FL} ahead illustrates this tradeoff at hand of a numerical example  in the special case $R_1=R_2=0$.) The tradeoff stems from the common random variable $U_0^0$ that is included in the exponent constraint \eqref{eq:thetaE3} for both $i\in\{1,2\}$, and from the rate-sharing of the  coding scheme in \cite{Michele_BC} for three different choices of $(\sigma_i,U_0^i,U_1^i,U_2^i)$, for $i=0,1,2$. 

	To see the effect of the expected rate-constraint in \eqref{eq:Ratei}, we compare above  exponents region $\mathcal{E}^*(R_0,R_1,R_2,\epsilon_1,\epsilon_2)$ with the exponents region $\mathcal{E}_{\text{fix}}^*(R_0,R_1,R_2,\epsilon_1,\epsilon_2)$ under more stringent maximum-length constraints 
	\begin{equation}\label{eq:maxRatei}
	\mathrm{len}\left(\M_i\right)\leq nR_i, \qquad i \in \{0,1,2\}.
	\end{equation} 
	In the limit $\epsilon_1, \epsilon_2 \downarrow 0$, the exponents region $\mathcal{E}_{\text{fix}}^*(R_0,R_1,R_2,\epsilon_1,\epsilon_2)$ was determined in \cite{Michele_BC}. Here, we strengthen this result by providing a strong converse, whose proof follows similar steps (but with the expected rate replaced by the maximum rate) as the converse to Theorem~\ref{thm1}.  
	\begin{thm}\label{thm2}
Under the maximum rate constraints \eqref{eq:maxRatei},   the exponents region $\mathcal{E}_{\text{fix}}^*(R_0,R_1,R_2,\epsilon_1,\epsilon_2)$ \emph{is independent of}  $(\epsilon_1,\epsilon_2)$ $\forall \, \epsilon_1 + \epsilon_2 <1$, and equals the set of 
 $(\theta_{1},\theta_{2})$ pairs satisfying:
\begin{subequations}\label{eq:E3_maxR}
	\begin{IEEEeqnarray}{rCl}
		\theta_{i} &\leq& I(U_0U_i;Y_i), \quad i \in \{1,2\}, \label{eq:theta2E3_maxR}\IEEEeqnarraynumspace
	\end{IEEEeqnarray}
	for some conditional pmfs $P_{U_0|Y_0}$, $P_{U_i|Y_0}$ satisfying
	\begin{IEEEeqnarray}{rCl}
		R_0& \geq & I(U_0;Y_0), \label{eq:R0une_maxR} \IEEEeqnarraynumspace\\
		R_i& \geq & I(U_i;Y_0|U_0), \quad i \in \{1,2\}.\label{eq:R2une_maxR}
	\end{IEEEeqnarray}	
\end{subequations}
	\end{thm}
\begin{IEEEproof}
Achievability is proved in \cite{Michele_BC}. The converse is proved in Appendix~\ref{strong_converse}.
\end{IEEEproof} 
Notice that \eqref{eq:E3_maxR} is obtained from \eqref{eq:E3} by setting $\sigma_0=1$ and $U_{0}^1, U_0^2, U_1^1, U_2^2$ constants. Moreover,  $\mathcal{E}_{\text{fix}}^*(R_0,R_1,R_2,\epsilon_1,\epsilon_2)=\mathcal{E}^*(R_0,R_1,R_2,0,0)$. Since $\mathcal{E}^*(R_0,R_1,R_2,\epsilon_1, \epsilon_2)$ is generally increasing in $(\epsilon_1,\epsilon_2)$,  expected rate-constraints allow to boost the exponents region compared to maximum rate-constraints.

\subsection{Only a Common Communication Link}
For $R_1=R_2=0$, i.e., without individual communication links, we can simplify the expression for $\mathcal{E}^*(R_0,R_1,R_2,\epsilon_1,\epsilon_2)$.
	\begin{definition}
		Define the two functions
			\begin{equation}
			\eta_i\left(R_0^i\right) := \max\limits_{\substack{P_{U_0^i|Y_0}\colon \\R_0^i \geq I\left(U_0^i;Y_0\right)}} I\left(U_0^i;Y_i\right), \qquad i \in \{1,2\},
		\end{equation}
		where the mutual information quantities are calculated with respect to the joint pmf $P_{U_0^iY_0Y_1Y_2} \triangleq P_{U_0^i|Y_0}P_{Y_0Y_1Y_2}$.
	\end{definition}

		  \begin{corollary}\label{cor}
	Let $\pi\colon \{1,2\}\to \{1,2\}$ be a permutation ordering the $\epsilon$-values in {decreasing} order:
	\begin{equation} 
		\epsilon_{\pi(1)} {\geq} \epsilon_{\pi(2) }.
	\end{equation}
	Then $\mathcal{E}^*(R_0,0,0,\epsilon_1,\epsilon_2)$ is the set of all ($\theta_{1},\theta_{2}$) pairs satisfying
		\vspace{-3mm}\begin{subequations}\label{eq:E3_simp2}
		\begin{IEEEeqnarray}{rCl}
			\theta_{\pi(1)} &\leq& I\left(U_0;Y_{\pi(1)}\right), \label{eq:theta1E3_simp2}\\
			\theta_{\pi(2)} &\leq& \min\big\{  I\left(U_0;Y_{\pi(2)}\right), \eta_{\pi(2)}\big(R_0^{\pi(2)}\big) \big\}, \label{eq:theta2E3_simp2}\IEEEeqnarraynumspace
		\end{IEEEeqnarray}
		for some conditional pmf $P_{U_0|Y_0}$ and rate $R_0^{\pi(2)}$ satisfying
		\begin{IEEEeqnarray}{rCl}
			R_0& \geq & \big(1-\epsilon_{\pi(1)}\big)I(U_0;Y_0) + \big(\epsilon_{\pi(1)}-\epsilon_{\pi(2) }\big)R_0^{\pi(2)}.\label{eq:R0une_simp2} \IEEEeqnarraynumspace
		\end{IEEEeqnarray}
	\end{subequations}
\end{corollary}
\begin{IEEEproof}
	See Appendix~\ref{app:Corollary}. 
\end{IEEEproof}

The following example illustrates the benefits of expected rate constraints versus maximum rate constraints, and the tradeoff between the two exponents when $R_1=R_2=0$. 
\begin{example}
Consider  the following joint pmf $P_{Y_0Y_1Y_2}$:

\vspace{2mm} 
\begin{tabular}{ |p{1.cm}| | p{1.25cm}|p{1.25cm}|p{1.25cm}|p{1.25cm}|  }
	\hline
	& \multicolumn{4}{|c|}{$(Y_1,Y_2)$} \\
	\hline 
	$Y_0$ & $(0,0)$ & $(0,1)$ & $(1,0)$ & $(1,1)$ \\
	\hline\hline
	$0$ & 0.05 & 0.05 & 0.15 & 0.083325\\
	\hline
	$1$ & 0.05 & 0.15 & 0.05 & 0.08335\\
	\hline
	$2$ & 0.15 & 0.05 & 0.05 & 0.083325\\
	\hline
\end{tabular}
\vspace{2mm}

For this pmf,  Figure \ref{fig:BC_Trenary_VL_FL} shows the optimal exponents regions under maximum- and expected-rate constraints when $R_0 =0.1$ and $\epsilon_1=0.15 >\epsilon_2=0.05$. The figure illustrates the boost in the exponents region due to the \emph{expected}-rate constraints.  It also emphasizes the benefits of sharing the rate in \eqref{eq:R0une_simp2} between two summands, which relate to the fact that depending on the observation $Y_0^n$ we use two variants of the coding scheme in \cite{Michele_BC}, one with auxiliary $U_0$ and the other with an auxiliary $U_0^{\pi(2)}$ that satisfies $I(U_0^{\pi(2)};Y_0)\leq R_0^{\pi(2)}$  and $I(U_0^{\pi(2)};Y_1)=\eta_{\pi(2)}(R_0^{\pi(2)})$. Restricting to a single auxiliary $U_0$ in  \eqref{eq:E3_simp2} (i.e., setting $R_0^{\pi(2)}= I(U_0;Y_0)$) results in an exponents region, denoted $\mathcal{E}_{\textnormal{no-RS}}\left(R_0,0,0,\epsilon_1,\epsilon_2\right)$ which coincides with $\mathcal{E}^*(R_0,0,0,\epsilon_2,\epsilon_2)$ and $\mathcal{E}_{\textnormal{fix}}^*\left((1-\epsilon_2)^{-1}R_0,0,0,\epsilon_1,\epsilon_2\right)$.
\begin{figure}[htbp]
	\begin{tikzpicture} [every pin/.style={fill=white},scale=.875]
		\begin{axis}[scale=.8,
			width=.93\columnwidth,
			scale only axis,
			xmin=0,
			xmax=0.011,
			xmajorgrids,
			xlabel={$\theta_{1}$},
			ymin=0,
			ymax=0.009,
			ymajorgrids,
			ylabel={$\theta_2$},
			axis x line*=bottom,
			axis y line*=left,
			legend style={at={(1.4,1)}, draw=none,fill=none,legend cell align=left, font=\normalsize}
			]

			\addplot[color=blue,dashdotted,line width=2pt]
			table[row sep=crcr]{		
				0.0 8.49746193e-03\\
				2.09823620e-03 8.49746193e-03\\
				0.00323431 0.00807244\\
 				0.00391825 0.00753612\\
                0.00522946 0.00636294\\
                0.00635515 0.00519916\\
                0.00740617 0.003942  \\
             	0.00794135 0.00325576\\
				8.46793368e-03 2.11467567e-03\\
				 8.46793368e-03 0\\};
			\addlegendentry{$\mathcal{E}_{\text{fix}}^*(R_0,0,0,\epsilon_1,\epsilon_2)$}
			
				\addplot[color=red,solid,dashed,line width=2pt]
			table[row sep=crcr]{
				0.         0.0089156 \\
				0.00220054 0.0089156 \\
				0.00363473 0.00827413\\
				0.00496227 0.00715514\\
				0.00675936 0.00546049\\
				0.00827377 0.00358815\\
				0.00842676 0.003371\\
				0.00888471 0.00230054\\
				0.00888471 0.        \\};
			\addlegendentry{$\mathcal{E}_{\text{no-RS}}(R_0,0,0,\epsilon_1,\epsilon_2)$}

			\addplot[color=teal,solid,line width=2pt]
			table[row sep=crcr]{
				 0.0 0.0089156 \\
 				 2.16225312e-03 0.0089156 \\
 				 2.22886173e-03 0.0089156 \\
				 3.44894205e-03 8.50447957e-03\\
				 4.26545664e-03 7.91515367e-03\\
				 5.45929809e-03 7.04159988e-03\\
				 6.89641461e-03 5.74229019e-03\\
				 8.25123317e-03 4.47311203e-03\\
				 9.98578868e-03 2.48940289e-03\\
				 1.00679017e-02 1.07276157e-03\\
				 1.00679017e-02 0.00000000e+00\\};
			\addlegendentry{$\mathcal{E}^*(R_0,0,0,\epsilon_1,\epsilon_2)$}
		\end{axis} 		
	\end{tikzpicture}
	\caption{Optimal error exponents regions under expected and maximum rate constraints  
		for $R_0=0.1,\epsilon_1=0.15,\epsilon_2=0.05$.}\vspace{-0.3cm}
	\label{fig:BC_Trenary_VL_FL} 
\end{figure}
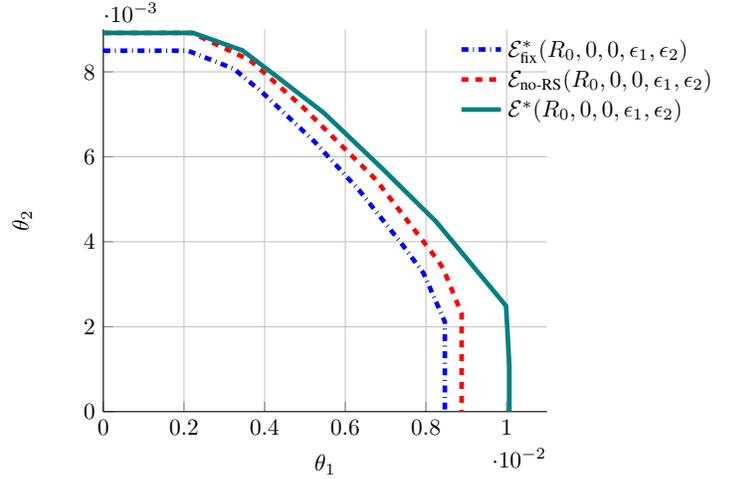

\end{example}

\section{Converse Proof to Theorem~\ref{thm1}}\label{general_converse}

Fix an exponent pair in $\mathcal{E}^*(R_0, R_1, R_2,\epsilon_1,\epsilon_2)$ and a sequence (in $n$) of encoding and decision functions $\{(\phi^{(n)}, g_1^{(n)}, g_2^{(n)})\}$ satisfying the constraints on the rate and the error probabilities in \eqref{eq:RPconstraints}. Our proof relies on the following lemma:\vspace{2mm}

\begin{lemma}\label{lem:receiverconverse}
	Fix a blocklength $n$ and a set $\mathcal{D}\subseteq \mathcal{Y}_0^n$ of positive probability, and let  the tuple ($\tilde{\M}_0,\tilde{\M}_1,\tilde{\M}_2,\tilde{Y}_0^n,\tilde{Y}_1^n,\tilde{Y}_2^n$)  follow the pmf 
	\begin{IEEEeqnarray}{rCl}
		\lefteqn{P_{{\tilde{\M}_0}{\tilde{\M}_1}{\tilde{\M}_2}\tilde{Y}_0^n\tilde{Y}_1^n\tilde{Y}_2^n}(\m_0,\m_1,\m_2,y_0^n,y_1^n,y_2^n) \triangleq} \qquad \qquad \nonumber \\
		 && P_{Y_0^nY_1^nY_2^n}(y_0^n,y_1^n,y_2^n)\cdot{\mathbbm{1} \{y_0^n\in \mathcal{D}\} \over P_{Y_0^n}(\mathcal{D})}\nonumber\\ && \qquad \cdot{\mathbbm{1}\{\phi^{(n)}(y_0^n)=(\m_0,\m_1,\m_2)\}}. \IEEEeqnarraynumspace \label{pmftildedoubleprime2_lemma}
	\end{IEEEeqnarray}
	Further, define ${U_0} \triangleq (\tilde{\M}_0,\tilde{Y}_0^{T-1},T)$, ${U_1} \triangleq \tilde{\M}_1$, $U_2 \triangleq \tilde{\M}_2$, $\tilde{Y}_i \triangleq\tilde{Y}_{i,T}$ (for $i\in\{0,1,2\}$),  where $T$ is uniform over $\{1,\ldots,n\}$ and independent of all other random variables. Notice the Markov chain $(U_0,U_1,U_2) \to \tilde{Y}_0 \to (\tilde{Y}_1,\tilde{Y}_2)$. Then the following inequalities hold:
	\begin{IEEEeqnarray}{rCl}
		H(\tilde{M}_0) &\geq& nI(U_0;\tilde{Y}_0) + \log P_{Y_0^n}(\mathcal{D}),\\
		H(\tilde{M}_i) &\geq& nI(U_i;\tilde{Y}_0|U_0), \quad i \in \{1,2\}.
	\end{IEEEeqnarray}
	Let $\eta > 0$ be arbitrary. For $i\in\{1,2\}$, if
	\begin{equation}\label{eq:lemma1_cond1}
		\Pr[\hat{\mathcal{H}}_{Y_i}=0|\mathcal{H}=0,Y_0^n=y_0^n]\geq \eta, \;\; \forall y_0^n \in \mathcal{D},
	\end{equation}
	then
	\begin{IEEEeqnarray}{rCl}
		-{1\over n}\log\beta_{i,n}
		&\leq& I(U_0U_i;\tilde{Y}_i) + \o_i(n),
	\end{IEEEeqnarray}
	where $\o_i(n)$ is a function that tends to $0$ as $n \to \infty$.
\end{lemma}
\begin{IEEEproof}
	See Appendix~\ref{app_lemma1}.
\end{IEEEproof}

We now proceed to prove the converse to Theorem~\ref{thm1}. Fix a positive $\eta>0$. Denote for each blocklength $n$, the set of strongly typical sequences in $\mathcal{Y}_0^n$ by $\mathcal{T}_{\mu_n}^{(n)}(P_{Y_0})$. Set $\mu_n=n^{-2/3}$ and define for $i\in\{1,2\}$, the sets 
\begin{IEEEeqnarray}{rCl}\label{Bn}
	\mathcal{B}_{i}(\eta) &\triangleq& \{y_0^n \in \mathcal{T}_{\mu_n}^{(n)}(P_{Y_0}) \colon \nonumber \\
	&& \; \mathrm{Pr}[\hat{\mathcal{H}}_{Y_i}=0 | Y_0^n = y_0^n, \mathcal{H}=0] \geq \eta\}, \; i\in\{1,2\},\IEEEeqnarraynumspace
	\label{eq:Bnidef}\\
	\mathcal{D}_{0}(\eta) &\triangleq& 	\mathcal{B}_{1}(\eta) \cap 	\mathcal{B}_{2}(\eta),\\
	\mathcal{D}_{i}(\eta) &\triangleq& \mathcal{B}_{i}(\eta) \backslash \mathcal{D}_{0}(\eta). \label{Din}
\end{IEEEeqnarray}
Further define for each $n$ the probabilities
\begin{IEEEeqnarray}{rCl}
	\Delta_j &\triangleq& P_{Y_0^n}(\mathcal{D}_{j}(\eta)), \quad j\in\{0,1,2\},
\end{IEEEeqnarray}
and notice that by the laws of probability
\begin{IEEEeqnarray}{rCl}\label{eq:sum}
	\Delta_0 +\Delta_i &=& P_{Y_0^n}(\mathcal{B}_{i}(\eta)), \quad i\in\{1,2\}, \\
	\Delta_0 & \geq & P_{Y_0^n}(\mathcal{B}_{1}(\eta)) + P_{Y_0^n}(\mathcal{B}_{2}(\eta))-1.
\end{IEEEeqnarray}
By   \eqref{type1constraint1}, it can be shown that
\begin{IEEEeqnarray}{rCl}
	1 - \epsilon_i  &\leq&
\eta(1- P_{Y_0^n}(\mathcal{B}_i(\eta))) + P_{Y_0^n}(\mathcal{B}_i(\eta)) +  P_{Y_0}^{n}(\overline{\mathcal{T}}_{\mu_n}^{(n)}).\IEEEeqnarraynumspace \label{eq:Bi_derivation} 
\end{IEEEeqnarray}
Thus, by \eqref{eq:Bi_derivation} and  \cite[Lemma~2.12]{Csiszarbook}:
\begin{IEEEeqnarray}{rCl}\label{eq:PB}
	P_{Y_0^n}(\mathcal{B}_{i}(\eta)) &\geq& {1 - \epsilon_i - \eta \over{1 - \eta}} - {\vert{\mathcal{Y}_0}\vert \over{(1-\eta)2 \mu_n n}}, \quad i\in\{1,2\}, \IEEEeqnarraynumspace
\end{IEEEeqnarray}
and we  conclude that in the limit $n\to \infty$ and $\eta \downarrow 0$:
\begin{subequations} \label{eq:Delta_conditions}
	\begin{IEEEeqnarray}{rCl}
		\lim_{\eta \downarrow 0} \lim_{n\to \infty}( \Delta_0+ \Delta_i )& \geq & 1- \epsilon_i, \quad i\in\{1,2\}\\
		\lim_{\eta \downarrow 0} \lim_{n\to \infty}\Delta_0 & \geq &  1- \epsilon_1 -\epsilon_2\\
		\lim_{\eta \downarrow 0} \lim_{n\to \infty}\sum_{j=0}^{2}\Delta_j & \leq &  1.
	\end{IEEEeqnarray}
\end{subequations}

We proceed by applying Lemma~\ref{lem:receiverconverse} to the set $\mathcal{D}_j$ for any $j\in\{0,1,2\}$ with $\Delta_j > 0$, and  conclude that for any $j\in \{0,1,2\}$ with  $\Delta_j > 0$ there is a tuple $(U_0^j,U_1^j,U_2^j)$ satisfying  
\begin{IEEEeqnarray}{rCl}
	H(\tilde{M}_{0}^{j}) &\geq& nI(U_0^j;\tilde{Y}_0^j) + \log P_{Y_0^n}(\mathcal{D}_j),\;\quad j\in\{0,1,2\}, \label{eq:M1i} \IEEEeqnarraynumspace\\
	H(\tilde{M}_{i}^{j}) &\geq& nI(U_{i}^{j};\tilde{Y}_{0}^j|U_0^{j}), \qquad i \in \{1,2\}, \; j\in\{0,i\},\IEEEeqnarraynumspace\label{eq:M2i}
\end{IEEEeqnarray}
and for $i \in \{1,2\},\; j\in\{0,i\}$:
\begin{IEEEeqnarray}{rCl}
	-\frac{1}{n} \log \beta_{i,n}
	&& \leq I(U_0^jU_i^j;\tilde{Y}_i^j) + \o_{i}^j(n),\IEEEeqnarraynumspace \label{eq:sumI}
\end{IEEEeqnarray}
where for each pair $(i,j)$,  the function $\o_{i}^{j}(n)\to 0$ as $n\to \infty$ and the random variables $\tilde{Y}_{0}^{j},\tilde{Y}_i^{j}, \tilde{M}_0^{j}, \tilde{M}_{i}^{j}$ are defined as in the lemma applied to the subset $\mathcal{D}_j$.

To summarize:
\begin{IEEEeqnarray}{rCl}
	-\frac{1}{n} \log \beta_{i,n}  &\leq& \min\{I(U_0^0U_i^0;\tilde{Y}_i^0);I(U_0^iU_i^i;\tilde{Y}_i^i)\} + \o_i(n),\IEEEeqnarraynumspace\label{eq:R1thetaslasteq}
\end{IEEEeqnarray}
where $\o_i(n)$ is a function tending to 0 as $n\to \infty$.

Define the following random variables for $i\in\{1,2\}$ and $j\in\{0,1,2\}$
\begin{equation}
	{\tilde{L}_{i,j}} \triangleq \mathrm{len}({\tilde{\M}_{i}^{j}}).
\end{equation}
By the rate constraints  \eqref{eq:Ratei}, and the definition of the random variables $\tilde{\M}_{i}^{j}$, we obtain by the total law of expectations
\begin{IEEEeqnarray}{rCl}
	nR_0 \geq \mathbb{E}[L_0] 
\geq\sum_{j\in\{0,1,2\}}\mathbb{E}[\tilde{L}_{0,j}]\Delta_j. \label{ELi}
\end{IEEEeqnarray} 
Moreover,  
\begin{IEEEeqnarray}{rCl}
	H(\tilde{\M}_{0}^{j}) &=& H(\tilde{\M}_{0}^{j},\tilde{L}_{0,j})\\
	&=& \sum_{l_j} \Pr[\tilde{L}_{0,j} = l_j]H(\tilde{\M}_{0}^{j}|\tilde{L}_{0,j}=l_j) + H(\tilde{L}_{0,j})\IEEEeqnarraynumspace\\
	&\leq& \sum_{l_j} \Pr[\tilde{L}_{0,j} = l_j]l_j + H(\tilde{L}_{0,j}) \label{HMi_ineq1}\\
	&= &\mathbb{E}[\tilde{L}_{0,j}] + H(\tilde{L}_{0,j}),
\end{IEEEeqnarray}
which combined with \eqref{ELi} establishes
\begin{IEEEeqnarray}{rCl}
	\hspace{-2mm}	\sum_{j\in\{0,1,2\}} \Delta_j H(\tilde{\M}_{0}^{j}) &\leq& \sum_{j\in\{0,1,2\}} \Delta_j\mathbb{E}[\tilde{L}_{0,j}] +  \Delta_j H(\tilde{L}_{0,j}) \IEEEeqnarraynumspace\\
	&\leq & {nR_0} \left(1 +\sum_{j\in\{0,1,2\}} h_b\left({\Delta_j \over nR_0}\right)\right), \label{M1ub'}\IEEEeqnarraynumspace
\end{IEEEeqnarray}
where \eqref{M1ub'} holds by \eqref{ELi} and because the entropy of a discrete and positive random variable $\tilde{L}_{0,j}$ of  mean   $\E{[\tilde{L}_{0,j}]} \leq {nR_0\over \Delta_j}$ is bounded by $ \frac{n R_0}{\Delta_j} \cdot h_b\left({\Delta_j \over nR_j}\right)$, see \cite[Theorem 12.1.1]{cover}. 

In a similar way we obtain for $i \in \{1,2\}$ 
\begin{IEEEeqnarray}{rCl}
	\sum_{j\in\{0,i\}}\Delta_j H(\tilde{\M}_{i}^{j}) \leq {nR_i} \left(1 +\sum_{j\in\{0,i\}} h_b\left({\Delta_j \over nR_i}\right)\right). \label{M2ub'} \IEEEeqnarraynumspace 
\end{IEEEeqnarray}

Notice that when  $\Delta_j = 0$, the trivial choice  $U_i^j=\tilde{Y}_i^j$ satisfies the inequalities \eqref{eq:R1thetaslasteq}, \eqref{M1ub'}, and \eqref{M2ub'}. Therefore, above conclusions  hold for $(U_0^j,U_1^j,U_2^j)$ for any $j\in\{0,1,2\}$.

Combining \eqref{M1ub'} and \eqref{M2ub'}  with \eqref{eq:M1i} and \eqref{eq:M2i}, noting \eqref{eq:sum} and \eqref{eq:PB}, and considering also \eqref{eq:R1thetaslasteq}, we have proved so far that for all $n\geq 1$ there exist joint pmfs $P_{U_0^jU_1^jU_2^j\tilde{Y}_0^j\tilde{Y}_1^j\tilde{Y}_2^j} = P_{\tilde{Y}_0^j}P_{\tilde{Y}_1^j\tilde{Y}_2^j|\tilde{Y}_0^j}P_{U_0^jU_1^jU_2^j|\tilde{Y}_0^j}$ (abbreviated as $P_{j}^{(n)}$) for $j\in\{0,1,2\}$ so that the following conditions hold for $i\in\{1,2\}$ (where $I_{P}$ indicates that the mutual information should be calculated according to a pmf $P$):
\begin{subequations}\label{eq:conditions}
	\begin{IEEEeqnarray}{rCl}			
		R_0 &\geq  &\sum_{j\in\{0,1,2\}}\big(I_{P_{j}^{(n)}}({U}_0^j;\tilde{Y}_0^j) + g_{1,j}(n)\big)\cdot g_{2,j}(n,\eta) ,\label{eq:R111} \IEEEeqnarraynumspace \\
		R_i &\geq  &\sum_{j\in\{0,i\}}\big(I_{P_{j}^{(n)}}({U}_i^j;\tilde{Y}_0^j|U_0^j)\big)\cdot g_{2,j}(n,\eta), \label{eq:R222}\\
		\theta_i &\leq& \min \{I_{P_{0}^{(n)}}({U}_0^0U_i^0;\tilde{Y}_i^0), I_{P_{i}^{(n)}}({U}_0^iU_i^i;\tilde{Y}_i^i)\} + g_{3,i}(n), \label{eq:theta111}\IEEEeqnarraynumspace
	\end{IEEEeqnarray}
\end{subequations}
for some nonnegative functions $g_{1,j}(n), g_{2,j}(n,\eta), g_{3,i}(n)$ with the following asymptotic behaviors:
\begin{IEEEeqnarray}{rCl}
	\lim_{n\to \infty} g_{1,j}(n) &=& 0, \qquad  \forall j \in \{0,1,2\},\IEEEeqnarraynumspace\\ 
	\lim_{n\to \infty} g_{3,i}(n) &=& 0, \qquad  \forall i \in \{1,2\},\\
	\lim_{n\to \infty} \left(g_{2,0}(n,\eta)+g_{2,i}(n,\eta)\right) & \geq  &\frac{1-\epsilon_i-\eta}{1-\eta}, \qquad \forall i \in \{1,2\}. \nonumber\\
\end{IEEEeqnarray}

By Carath\'eodory's theorem \cite[Appendix C]{ElGamal}, there exist for each $n$, random variables ${U}_0^0,U_0^1,U_0^2,U_1^0,U_1^1,U_2^0,U_2^2$ satisfying \eqref{eq:conditions} over alphabets of sizes
\begin{align}
	\vert {\mathcal{U}}_0^0\vert &\leq \vert \mathcal{Y}_0\vert + 3, \\
	\vert {\mathcal{U}}_0^j\vert &\leq \vert \mathcal{Y}_0\vert + 2, \qquad \quad j \in \{1,2\},\\
	\vert {\mathcal{U}}_i^j \vert &\leq \vert {\mathcal{U}}_0^j \vert\cdot|\mathcal{Y}_0| + 1, \quad i \in \{1,2\}, j \in \{0,i\}.
\end{align}
Then we invoke the Bolzano-Weierstrass theorem and  consider for each $j \in \{0,1,2\}$ a sub-sequence  $P_{U_0^jU_1^jU_2^j\tilde{Y}_0^j\tilde{Y}_1^j\tilde{Y}_2^j}^{(n_k)}$ that converges to a limiting pmf $P_{U_0^jU_1^jU_2^jY_0^jY_1^jY_2^j}^{*}$. For these limiting pmfs, which we abbreviate by $P_{j}^*$, we conclude by \eqref{eq:R111}--\eqref{eq:theta111} and \eqref{eq:Delta_conditions} that for all $i\in\{1,2\}$:
\begin{IEEEeqnarray}{rCl}			
	R_0& \geq & \sigma_0\cdot  I_{P_{0}^{*}}({U}_0^0;{Y}_0^0) + \sigma_1 \cdot I_{P_{1}^{*}}({U}_0^1;{Y}_0^1)\nonumber\\
	&& \quad +  \sigma_2 \cdot I_{P_{2}^{*}}({U}_0^2;{Y}_0^2),\label{eq:R_1_f} \\
	R_i &\geq  &\sigma_0\cdot  I_{P_{0}^{*}}({U}_i^0;{Y}_0^0|U_0^0) +   \sigma_i \cdot I_{P_{i}^{*}}({U}_i^i;{Y}_0^i|U_0^i),\\
	\theta_i &\leq & \min \{I_{P_{0}^{*}}(U_0^0{U}_i^0;{Y}_i^0),I_{P_{i}^{*}}(U_0^i{U}_i^i;{Y}_i^i)\}, \label{theta_1_f}
\end{IEEEeqnarray}
where numbers $\sigma_0,\sigma_1,\sigma_2 >0$ satisfy $\sigma_0+\sigma_1+\sigma_2 \leq 1$ and 
\begin{subequations}\label{eq:cond_epsilon}
	\begin{IEEEeqnarray}{rCl}\label{eq:sigma_constraints_KHop}
		\sigma_{0} +\sigma_i  & \geq &   1-\epsilon_i, \qquad i\in \{1,2\},\\
		\sigma_0 & \geq & 1-\epsilon_1-\epsilon_2.
	\end{IEEEeqnarray}
\end{subequations}
Notice further that since for any $j \in\{0,1,2\}$ and any $k$, the sequence $\tilde{Y}_0^{j,n_k}$ lies in the typical set $\mathcal{T}^{(n_k)}_{\mu_{n_k}}(P_{Y_0})$, we have  for all $j \in \{0,1,2\}$, $\vert P_{\tilde{Y}_0^j} - P_{Y_0}\vert \leq \mu_{n_k}$ and thus the limiting pmf satisfies $P^*_{Y_0^j}=P_{Y_0}$. Moreover, since for each $n_k$ the pair of random  variables $(\tilde{Y}_1^j,\tilde{Y}_2^j)$ is drawn according to $P_{Y_1Y_2|Y_0}$ given $\tilde{Y}_0^j$, the limiting pmf also satisfies $P_{Y_1^jY_2^j|Y_0^j}^*=P_{Y_1Y_2|Y_0}$. 
We also notice for all $j \in \{0,1,2\}$ that  under $P_{j}^*$ the  Markov chain
$(U_0^j,U_1^j,U_2^j)\to Y_0 \to (Y_1,Y_2)$ holds. 
This concludes the converse proof.

	\appendices

	\section*{Acknowledgment}
M. Wigger and M. Hamad have been supported by the European Union’s Horizon 2020 Research And Innovation Programme under grant agreement no. 715111.

\section{Achievability proof for Theorem~\ref{thm1}}\label{app:Ach.}
\subsection{The Scheme}
Choose random variables $U_0^0, U_1^0, U_2^0, U_0^1, U_1^1,U_0^2, U_2^2$ and probabilities $\sigma_{0},\sigma_{1},\sigma_{2}$, so that  $\sigma_{0} + \sigma_{1} + \sigma_{2} \leq 1$, and Conditions~\eqref{eq:R0une}--\eqref{eq:sigmas2} are satisfied. 
Define three disjoint sets $\mathcal{D}_0,\mathcal{D}_1,\mathcal{D}_2\subseteq \mathcal{Y}_0^n$ with  probabilities (under $P_{Y_0}^n$) equal to $\sigma_{0},\sigma_{1},\sigma_{2}$, respectively. Let  $\mathcal{S}:=\mathcal{Y}_0^{n}\backslash (\mathcal{D}_0,\mathcal{D}_1,\mathcal{D}_2)$ denote their complement, which has probability $1-(\sigma_{0}+\sigma_{1}+\sigma_{2})$. 

Whenever $Y_0^n \in \mathcal{S}$, the transmitter T$_{Y_0}$ sends the two-bit message
\begin{equation}
\M_0 = [0,0]
\end{equation}
over the common link and nothing over the individual links $\M_1=\M_2=\emptyset$. 
Upon receiving these messages, both decision centers R$_{Y_1}$ and R$_{Y_2}$ decide on
\begin{equation}
\hat{\mathcal{H}}_{Y_1} = \hat{\mathcal{H}}_{Y_2} =  1. 
\end{equation}

Whenever $Y_0^n \in \mathcal{D}_0$, then T$_{Y_0}$, R$_{Y_1}$, R$_{Y_2}$ all follow the  coding scheme in \cite{Michele_BC} with the choice of auxiliaries $U_0^0, U_1^0, U_2^0$. 
Additionally, T$_{Y_0}$ adds [0,1]-flag bits to the common messages $\M_0$ to indicate to R$_{Y_1}$ and R$_{Y_2}$ that $Y_0^n \in \mathcal{D}_0$.

Whenever $Y_0^n \in \mathcal{D}_1$, then T$_{Y_0}$ and  R$_{Y_1}$ follow the  coding scheme  in \cite{Michele_BC} with the choice of auxiliaries $U_0^1, U_1^1$. 
Additionally, T$_{Y_0}$ adds [1,0]-flag bits to its common messages $\M_0$ to indicate to R$_{Y_1}$ and R$_{Y_2}$ that $Y_0^n \in \mathcal{D}_1$. Note that no message is sent over the individual link to R$_{Y_2}$, i.e., $\M_2=\emptyset$. Moreover, 
R$_{Y_2}$ declares $\hat{\mathcal{H}}_{Y_2} = 1$.

Whenever $Y_0^n \in \mathcal{D}_2$, then  T$_{Y_0}$ and R$_{Y_2}$ follow the  coding scheme  in \cite{Michele_BC} with the choice of auxiliaries $U_0^2, U_2^2$.
Additionally, T$_{Y_0}$ adds [1,1]-flag bits to its common messages $\M_0$ to indicate to R$_{Y_1}$ and R$_{Y_2}$ that $Y_0^n \in \mathcal{D}_2$. Note that no message is sent over the individual link to R$_{Y_1}$, i.e., $\M_1=\emptyset$. 
Moreover,  R$_{Y_1}$  declares $\hat{\mathcal{H}}_{Y_1} = 1$.

\subsection{Analysis}

Let ${\tilde{\mathcal{H}}}_{Y_i}^{(j)}$ denote the hypothesis guessed by {{R$_{Y_i}$}}, for $i\in\{1,2\}$, and $\tilde{R}_i^{(j)}$ the required rate of message $M_i$,  for $i\in\{0,1,2\}$,   when the scheme in \cite{Michele_BC} is employed  with auxiliaries $(U_0^j,U_1^j,U_2^j)$, for $j\in\{0,1,2\}$, to the present setup.  We can then write:
\begin{IEEEeqnarray}{rCl}
	\alpha_{1,n} &=& \Pr[\hat{\mathcal{H}}_{Y_1} = 1| {\mathcal{H}} = 0] \\
	&=& \Pr[\hat{\mathcal{H}}_{Y_1} = 1, Y_0^n \in \mathcal{S} |\mathcal{H} = 0] \nonumber \\
	&& + \Pr[\hat{\mathcal{H}}_{Y_1} = 1, Y_0^n \in (\mathcal{D}_{0}\cup \mathcal{D}_1) |\mathcal{H} = 0] \nonumber \\
	&& + \Pr[\hat{\mathcal{H}}_{Y_1} = 1, Y_0^n \in \mathcal{D}_{2}|\mathcal{H} = 0] \IEEEeqnarraynumspace\\
	&=& \Pr[Y_0^n \in \mathcal{S}] + \Pr[Y_0^n \in \mathcal{D}_{2}] \nonumber \\ 
	&& + \Pr[\tilde{\mathcal{H}}_{Y_1}^{(0)} = 1, Y_0^n \in \mathcal{D}_{0} |\mathcal{H} = 0] \IEEEeqnarraynumspace \nonumber\\
	&& + \Pr[\tilde{\mathcal{H}}_{Y_1}^{(1)} = 1, Y_0^n \in \mathcal{D}_{1} |\mathcal{H} = 0] \IEEEeqnarraynumspace\\    
	&\leq& \epsilon_1 + \Pr[\tilde{\mathcal{H}}_{Y_1}^{(0)} = 1 |\mathcal{H} = 0]  + \Pr[\tilde{\mathcal{H}}_{Y_1}^{(1)} = 1 |\mathcal{H} = 0], \IEEEeqnarraynumspace
\end{IEEEeqnarray}
because $\Pr[Y_0^n \in \mathcal{S}] + \Pr[Y_0^n \in \mathcal{D}_{2}] = 1- \sigma_0 - \sigma_1 \leq \epsilon_1$ by \eqref{eq:sigmas}. Analogously, we have 
\begin{IEEEeqnarray}{rCl}
	\alpha_{2,n} 
	&\leq& \epsilon_2 + \Pr[\tilde{\mathcal{H}}_{Y_2}^{(0)} = 1 |\mathcal{H} = 0] + \Pr[\tilde{\mathcal{H}}_{Y_2}^{(2)} = 1 |\mathcal{H} = 0] .\IEEEeqnarraynumspace
\end{IEEEeqnarray}
Since by \cite{Michele_BC}, $\Pr[\tilde{\mathcal{H}}_{Y_1}^{(0)} = 1 |\mathcal{H} = 0]$, $\Pr[\tilde{\mathcal{H}}_{Y_1}^{(1)} = 1 |\mathcal{H} = 0]$, $\Pr[\tilde{\mathcal{H}}_{Y_2}^{(0)} = 1 |\mathcal{H} = 0]$   and $\Pr[\tilde{\mathcal{H}}_{Y_2}^{(2)} = 1 |\mathcal{H} = 0]$ all tend to 0 as $n \to \infty$, we conclude that for the above coding scheme, $\varlimsup_{n\to\infty}\alpha_{1,n} \leq \epsilon_1$ and $\varlimsup_{n\to\infty}\alpha_{2,n} \leq \epsilon_2$. 

\vspace{2mm}
For the type-II error probabilities we obtain
\begin{IEEEeqnarray}{rCl}
	\beta_{1,n} &=& \Pr[\hat{\mathcal{H}}_{Y_1} = 1| {\mathcal{H}} = 0] \\
	&=& \Pr[\hat{\mathcal{H}}_{Y_1} = 1, Y_0^n \in \mathcal{S} |\mathcal{H} = 0] \nonumber \\
	&& + \Pr[\hat{\mathcal{H}}_{Y_1} = 1, Y_0^n \in \mathcal{D}_{0} |\mathcal{H} = 0] \nonumber \\
	&& + \Pr[\hat{\mathcal{H}}_{Y_1} = 1, Y_0^n \in \mathcal{D}_{1} |\mathcal{H} = 0] \nonumber \\
	&& + \Pr[\hat{\mathcal{H}}_{Y_1} = 1, Y_0^n \in \mathcal{D}_{2}|\mathcal{H} = 0] \IEEEeqnarraynumspace\\
	&=& \Pr[\tilde{\mathcal{H}}_{Y_1}^{(0)} = 1, Y_0^n \in \mathcal{D}_{0} |\mathcal{H} = 0] \IEEEeqnarraynumspace \nonumber \\
	&& + \Pr[\tilde{\mathcal{H}}_{Y_1}^{(1)} = 1, Y_0^n \in \mathcal{D}_{1} |\mathcal{H} = 0] \\  
	&\leq& \Pr[\tilde{\mathcal{H}}_{Y_1}^{(0)} = 1 |\mathcal{H} = 0] + \Pr[\tilde{\mathcal{H}}_{Y_1}^{(1)} = 1 |\mathcal{H} = 0]
\end{IEEEeqnarray}
and analogously
\begin{IEEEeqnarray}{rCl}
	\beta_{2,n} 
	&\leq& \Pr[\tilde{\mathcal{H}}_{Y_2}^{(0)} = 1 |\mathcal{H} = 0] + \Pr[\tilde{\mathcal{H}}_{Y_2}^{(2)} = 1 |\mathcal{H} = 0]. 
\end{IEEEeqnarray}
Taking logarithms, dividing by the blocklength $n$, and letting $n\to \infty$, we then obtain for $i\in\{1,2\}$:
\begin{IEEEeqnarray}{rCl}
\lefteqn{
\varliminf_{n\to \infty} - \frac{1}{n} \log \beta_{i,n}  } \nonumber \\
& = &  \min \Big\{ \varliminf_{n\to \infty} - \frac{1}{n} \log\Pr[\tilde{\mathcal{H}}_{Y_i}^{(0)} = 1 |\mathcal{H} = 0]  , \nonumber \\
& & \qquad\; \qquad \varliminf_{n\to \infty} - \frac{1}{n} \log\Pr[\tilde{\mathcal{H}}_{Y_i}^{(i)} = 1 |\mathcal{H} = 0] \Big\}\\
& = & \min\{ I(U_0^0U_i^0;Y_i), \; I(U_0^iU_i^i;Y_i)\}, 
\end{IEEEeqnarray}
where the last equality holds  by \cite{Michele_BC}.

\vspace{2mm}
Finally,  the expected lengths of the messages are given by
\begin{IEEEeqnarray}{rCl}
	\E[\len(\M_0)] &\leq& 2 + \sum_{j\in\{0,1,2\}}  \sigma_{j} \cdot  n \tilde{R}_0^{(j)}   
\end{IEEEeqnarray}
and for $i\in\{1,2\}$ 
\begin{IEEEeqnarray}{rCl}
	\E[\len(\M_i)] &\leq& \sigma_{0} n \tilde{R}_i^{(0)} + \sigma_{i} n \tilde{R}_i^{(i)}. 
\end{IEEEeqnarray}
Since the chosen random variables  $U_0^0, U_1^0, U_2^0, U_0^1, U_1^1,U_0^2, U_2^2$ and probabilities $\sigma_{0},\sigma_{1},\sigma_{2}$ satisfy Conditions~\eqref{eq:R0une}--\eqref{eq:R2une},  and since by \cite{Michele_BC} , for $i \in \{1,2\}$ and $j \in \{0,1,2\}$
\begin{IEEEeqnarray}{rCl}
 \tilde{R}_0^{(j)} &=& I(U_0^{j}; Y_0)+ \mu, \\
 \tilde{R}_i^{(0)} &=& I(U_i^{0}; Y_0| U_0^{0})+ \mu, \\
 \tilde{R}_i^{(i)} &=& I(U_i^{i}; Y_0|U_0^{i})+ \mu,
\end{IEEEeqnarray} for an arbitrary small $\mu>0$, we conclude that in the limit $n \to \infty$ and $\mu\downarrow 0$ the expected lengths of the messages satisfy the rate constraints \eqref{eq:Ratei}.

\section{Proof of Lemma~\ref{lem:receiverconverse}}\label{app_lemma1}
Throughout this section, let $h_{b}(\cdot)$ denote the binary entropy function, and $D(P\|Q)$  the Kullback-Leibler divergence between two probability mass functions on the same alphabet.
Note first that by \eqref{pmftildedoubleprime2_lemma}:
\begin{equation}\label{tildedivergencerelation2_lemma}
	D(P_{\tilde{Y}_0^n}\|P_{Y_0}^{n}) \leq \log{\Delta_n^{-1}},
\end{equation}
where we defined $\Delta_n \triangleq P_{Y_0^n}(\mathcal{D})$.

Further define  $\tilde{U}_{0,t}\triangleq(\tilde{\M}_0,\tilde{Y}_0^{t-1})$ and $\tilde{U}_{1,t}\triangleq\tilde{\M}_1$, $\tilde{U}_{2,t}\triangleq\tilde{\M}_2$ and notice:
\begin{IEEEeqnarray}{rCl}
	H(\tilde{\M}_0)	&\geq& I(\tilde{\M}_0;\tilde{Y}_0^n) + D(P_{\tilde{Y}_0^n}\|P_{Y_0}^n) + \log\Delta_{n}\IEEEeqnarraynumspace\label{m1entropylbstep1_lemma}\\
	&=& H(\tilde{Y}_0^n) + D(P_{\tilde{Y}_0^n}\|P_{Y_0}^n) \nonumber \\ && -  H(\tilde{Y}_0^n|\tilde{\M}_0) + \log\Delta_{n}\IEEEeqnarraynumspace\\
	&\geq& n [H(\tilde{Y}_{0,T}) + D(P_{\tilde{Y}_{0,T}}\|P_{Y_0})] \nonumber \\ && - \sum_{t=1}^{n} H(\tilde{Y}_{0,t}|\tilde{U}_{0,t})+ \log\Delta_{n}\IEEEeqnarraynumspace\label{m1entropylbstep4_lemma}\\
	&=& n [H(\tilde{Y}_{0,T}) + D(P_{\tilde{Y}_{0,T}}\|P_{Y_0}) \nonumber \\ && - H(\tilde{Y}_{0,T}|\tilde{U}_{0,T},T)]+ \log\Delta_{n} \label{Tuniformdef_lemma}\\
	&\geq & n [H(\tilde{Y}_{0,T})  - H(\tilde{Y}_{0,T}|\tilde{U}_{0,T},T)]+ \log\Delta_{n} \IEEEeqnarraynumspace\\
	&=& n [I(\tilde{Y}_0;U_0) + {1 \over n}  \log{\Delta_{n}}] .\IEEEeqnarraynumspace \label{eq:HM1_LB_lemma_last_eq}
\end{IEEEeqnarray}
Here, (\ref{m1entropylbstep1_lemma}) holds by (\ref{tildedivergencerelation2_lemma}); (\ref{m1entropylbstep4_lemma}) holds by the super-additivity property in \cite[Proposition 1]{tyagi2019strong}, by the chain rule, and by the definition of $\tilde{U}_{0,t}$; \eqref{Tuniformdef_lemma} by defining $T$ uniform over $\{1,\dots,n\}$ independent of all other random variables; and \eqref{eq:HM1_LB_lemma_last_eq} by the definitions of $U_0$ and $\tilde{Y}_0$ in the lemma.

We  lower bound the entropy of $\tilde{\M}_1$ and $\tilde{\M}_2$  for $i\in\{1,2\}$:
\begin{IEEEeqnarray}{rCl}
	H(\tilde{\M}_i) &\geq& I(\tilde{\M}_i;\tilde{Y}_0^n|\tilde{\M}_0) \label{eq:Mi_functionof_Y0}\\
	&\geq& \sum_{t=1}^{n} I(\tilde{\M}_i;\tilde{Y}_{0,t}|\tilde{\M}_0\tilde{Y}_0^{t-1}) \\
	&=&	n I(U_i;\tilde{Y}_{0,T}|\tilde{U}_{0,T},T)\label{eq:def_Ui_lemma}\\
	&=& n I(U_i;\tilde{Y}_{0}|{U}_{0})\label{eq:def_Y0_lemma} 
\end{IEEEeqnarray}
where \eqref{eq:Mi_functionof_Y0} holds since conditioning can only  reduce entropy and since $\tilde{\M}_i$ is a function of $\tilde{Y}_0^n$, and \eqref{eq:def_Ui_lemma}--\eqref{eq:def_Y0_lemma} hold by the definitions of $\tilde{U}_{0,T}$, $U_1$, $U_2$, $\tilde{Y}_0$, and $U_0$.

We next  upper bound the error exponents at the decision centers. In the following, we note that the pair $(\m_0,\m_i)$ is always determined as a function of $y_0^n$. 

Define for $i\in\{1,2\}$
\begin{equation}\label{eq:yz_acceptance}
	\mathcal{A}_{Y_i,n}(\m_0,\m_i) \triangleq \{y_i^n \colon g_i(\m_0,\m_i,y_i^n) = 0\},
\end{equation}
and its Hamming neighborhood:
\begin{IEEEeqnarray}{rCl}
	\hat{\mathcal{A}}_{Y_i,n}^{\ell_n}(\m_0,\m_i) &\triangleq& \{\tilde{y}_i^n : \exists \, y_i^n \in \mathcal{A}_{Y_i,n}(\m_0,\m_i) \nonumber \\
	&& \qquad \qquad \textnormal{ s.t.} \; d_H(y_i^n,\tilde{y}_i^n)\leq\ell_n\}
\end{IEEEeqnarray}
for some real number $\ell_n$ satisfying $\lim_{n \rightarrow \infty} {\ell_n/n} =0 $ and $\lim_{n \to \infty} {\ell_n/\sqrt{n}} =\infty $.	
Since by Condition \eqref{eq:lemma1_cond1}, 
\begin{equation}\label{blowupcond2_lemma}
	P_{\tilde{Y}_i^n|\tilde{Y}_0^n}(\mathcal{A}_{Y_i,n}(\m_0,\m_i)|y_0^n) \geq \eta , \quad \forall y_0^n \in \mathcal{D},
\end{equation}
by the blowing-up lemma \cite{MartonBU}:
\begin{equation}\label{blowup2_lemma}
	P_{\tilde{Y}_i^n|\tilde{Y}_0^n}\left(\hat{\mathcal{A}}_{Y_i,n}^{\ell_n}(\m_0,\m_i)|y_0^n\right) \geq 1 - \zeta_n, \quad \forall y_0^n \in \mathcal{D},
\end{equation}
for a real number $\zeta_n > 0$ such that $\lim\limits_{n \to \infty} \zeta_n = 0$.\\
Define
\begin{equation}
	{\mathcal{A}}_{Y_i,n} \triangleq \bigcup\limits_{(\m_0,\m_i)\in \mathcal{M}_0\times\mathcal{M}_i} \{\m_0,\m_i\} \times {\mathcal{A}}_{Y_i,n}(\m_0,\m_i)
\end{equation}
and
\begin{equation}
	\hat{\mathcal{A}}_{Y_i,n}^{\ell_n} \triangleq \bigcup\limits_{(\m_0,\m_i )\in \mathcal{M}_0\times\mathcal{M}_i} \{\m_0,\m_i\} \times \hat{\mathcal{A}}_{Y_i,n}^{\ell_n}(\m_0,\m_i),
\end{equation} 
and notice that
\begin{IEEEeqnarray}{rCl}
	\lefteqn{P_{\tilde{\M}_0\tilde{\M}_i\tilde{Y}_i^n}(\hat{\mathcal{A}}_{Y_i,n}^{\ell_n}) } \quad \nonumber \\
	&= & \sum_{ y_0^n\in\mathcal{D} } \; P_{\tilde{Y}_0^n}(y_0^n) \cdot P_{\tilde{Y}_i^n|\tilde{Y}_0^n}({\hat{\mathcal{A}}}_{Y_i,n}^{\ell_n}(\m_0,\m_i)|y_0^n)\IEEEeqnarraynumspace\\
	&\geq&  (1-\zeta_n).\label{eq:zetan}
\end{IEEEeqnarray}
Finally, we can write
\begin{IEEEeqnarray}{rCl}
	\lefteqn{P_{\tilde{\M}_0\tilde{\M}_i}P_{\tilde{Y}_i^n}\left(\hat{\mathcal{A}}_{Y_i,n}^{\ell_n}\right)}\quad \,\nonumber\\
	&\leq& P_{\M_0\M_i}P_{Y_i}^n\left(\hat{\mathcal{A}}_{Y_i,n}^{\ell_n}\right)\Delta_n^{-2}\\
	& = & \sum_{ \substack{(\m_0,\m_i) \in \\ \mathcal{M}_0\times \mathcal{M}_i}} P_{\M_0\M_i} (\m_0,\m_i)P_{{Y}_i}^n\left( \hat{\mathcal{A}}_{Y_i,n}^{\ell_n}(\m_0,\m_i)\right)\Delta_n^{-2}\IEEEeqnarraynumspace\\
	&\leq& \sum_{ \substack{(\m_0,\m_i) \in\\ \mathcal{M}_0\times \mathcal{M}_i}} P_{\M_0\M_i} (\m_0,\m_i) P_{{Y}_i}^n\left( {\mathcal{A}}_{Y_i,n}(\m_0,\m_i)\right) \nonumber \\
	& & \hspace{1cm} \cdot e^{nh_b(\ell_n/n)}|\mathcal{Y}_i|^{\ell_n}k_n^{\ell_n}\Delta_{n}^{-2}\IEEEeqnarraynumspace\\
	&=& \beta_{i,n} e^{n \delta_n},\label{Eq:ByCsiszarKornerLemma_lemma}
\end{IEEEeqnarray}
where  $\delta_n\triangleq h_b(\ell_n/n) + \frac{\ell_n}{n} \log ( |\mathcal{Y}_i|\cdot k_n) -\frac{2}{n}\log \Delta_n $ and $k_n \triangleq \min\limits_{\substack{y_i,y_i':\\P_{Y_i}(y_i') > 0}}{P_{Y_i}(y_i) \over P_{Y_i}(y_i')}$.
Here, (\ref{Eq:ByCsiszarKornerLemma_lemma}) holds by \cite[Proof of Lemma 5.1]{Csiszarbook}.

Combining \eqref{Eq:ByCsiszarKornerLemma_lemma} with  \eqref{eq:zetan} and standard inequalities (see \cite[Lemma~1]{JSAIT}), we  then obtain:\vspace{2mm}
\begin{IEEEeqnarray}{rCl}\label{theta_ub_lemma}
	\lefteqn{-{1\over n}\log \beta_{i,n}} \qquad \nonumber\\
	& \leq & 	-{1\over n}\log \left( P_{\tilde{\M}_0\tilde{\M}_i}P_{\tilde{Y}_i^n}\left(\hat{\mathcal{A}}_{Y_i,n}^{\ell_n}\right) \right) +  \delta_n\\
	&\leq& {1 \over n (1-\zeta_n)} D(P_{\tilde{\M}_0\tilde{\M}_i\tilde{Y}_i^n}\|P_{\tilde{\M}_0\tilde{\M}_i}P_{\tilde{Y}_i^n}) + \delta_n +\frac{1}{n},\IEEEeqnarraynumspace
\end{IEEEeqnarray}
where $\zeta_n$ and $\delta_n$ both tend to 0 as $n \to \infty$.
We continue to upper bound the divergence term as
\begin{IEEEeqnarray}{rCl}
	{D(P_{\tilde{\M}_0\tilde{\M}_i\tilde{Y}_i^n}\|P_{\tilde{\M}_0\tilde{\M}_i}P_{\tilde{Y}_i^n})}
	&=& I(\tilde{\M}_0\tilde{\M}_i;\tilde{Y}_i^n) \\
	&=& \sum_{t=1}^n I(\tilde{\M}_0\tilde{\M}_i;\tilde{Y}_{i,t}|\tilde{Y}_i^{t-1}) \label{eq:divergence_chainrule}\\
	&\leq& \sum_{t=1}^n I(\tilde{\M}_0\tilde{\M}_i\tilde{Y}_0^{t-1};\tilde{Y}_{i,t}) \IEEEeqnarraynumspace\label{eq:divergence_markovcahins}\\
	&=& n[I(\tilde{U}_{0,T}U_i;\tilde{Y}_{i,T}|T)]\label{eq:by_definitions_start}\\
	&\leq& n[I(\tilde{U}_{0,T}TU_i;\tilde{Y}_{i,T})]\\
	&=& n [I(U_0U_i;\tilde{Y}_i) ]\label{theta_ub2_lemma}.
\end{IEEEeqnarray}
Here, \eqref{eq:divergence_chainrule} holds by the chain rule; \eqref{eq:divergence_markovcahins} by the Markov chain $\tilde{Y}_i^{t-1} \to (\tilde{Y}_0^{t-1}\tilde{M}_0\tilde{M}_i) \to \tilde{Y}_{i,t}$; and \eqref{eq:by_definitions_start}--\eqref{theta_ub2_lemma} by the definitions of $T, \tilde{U}_{0,t},U_0,U_i,\tilde{Y}_i$.

\section{Proof of Corollary~\ref{cor}}
 \label{app:Corollary}
By  Theorem~\ref{thm1}, $\mathcal{E}^*(R_0,0,0, \epsilon_1,\epsilon_2)$ is 
the set of all ($\theta_{1},\theta_{2}$) pairs  satisfying 
\begin{subequations}\label{eq:E3_simp1}
	\begin{equation}
		\theta_{i} \leq \min\big\{ I(U_0;Y_i), \eta_i\big(R_0^i\big)\big\},  \qquad i \in \{1,2\}. \label{eq:thetaE3_simp1}
	\end{equation}
	for some non-negative numbers $\sigma_0,\sigma_{1},\sigma_2$  with sum $\leq 1$ and satisfying \eqref{eq:sigmas} and \eqref{eq:sigmas2}, a conditional pmf $P_{U_0|Y_0}$, and nonnegative rates $R_0^1,R_0^2$ such that
	\begin{IEEEeqnarray}{rCl}
		R_0& \geq & \sigma_{0}I(U_0;Y_0) + \sigma_{1} R_0^1 + \sigma_{2}R_0^2. \label{eq:R0une_simp1} \IEEEeqnarraynumspace 
	\end{IEEEeqnarray}
\end{subequations}
Notice  that without loss in optimality,  in the evaluation of above region, we can restrict to tuples  $\big(P_{U_0|Y_0},R_0^1,R_0^2\big)$ satisfying
\begin{equation}
I(U_0;Y_i) \geq \eta_{i}\left(R_0^{i}\right) ,\label{eq:initial_assumption}
\end{equation}
which by the maximum in the definition of function $\eta_{i}$ implies 
\begin{equation}
 I(U_0;Y_0) \geq R_0^{i}, \quad i \in \{1,2\}.\label{eq:initial_assumption_R} 
\end{equation}
In fact, if \eqref{eq:initial_assumption} is violated,  rates $R_0^{1}$ and/or $R_0^{2}$  can be reduced without  changing  \eqref{eq:thetaE3_simp1} and so that \eqref{eq:initial_assumption}  holds.

We  next show that  any exponent pair $(\theta_1,\theta_2)$ and tuple $(P_{U_0|Y_0}, R_0^1, R_0^2)$ satisfying \eqref{eq:E3_simp1}, \eqref{eq:initial_assumption}, and 
\begin{equation}\label{eq:condU}
I(U_0;Y_0) \leq R_0^{\pi(1)} + R_0^{\pi(2)}
\end{equation} also satisfies \eqref{eq:E3_simp2}. The exponents' constraints \eqref{eq:theta1E3_simp2} and \eqref{eq:theta2E3_simp2}  are easily verified. 
To verify \eqref{eq:R0une_simp2}, notice that when $\sigma_{0} > 1-\epsilon_{\pi(1)}$:
\begin{IEEEeqnarray}{rCl}
	R_0 &\geq& \sigma_{0}I(U_0;Y_0) + \sigma_{\pi(1)}R_0^{\pi(1)} + \sigma_{\pi(2)}R_0^{\pi(2)} \\
	&=& (1-\epsilon_{\pi(1)}) I(U_0;Y_0) + \sigma_{\pi(1)}R_0^{\pi(1)}  \nonumber \\ && + (\sigma_{0}-1+\epsilon_{\pi(1)})I(U_0;Y_0)
	+ \sigma_{\pi(2)}R_0^{\pi(2)} \label{eq:holds_by_case2}\\
	&\geq& (1-\epsilon_{\pi(1)}) I(U_0;Y_0) + (\epsilon_{\pi(1)}-\epsilon_{\pi(2)})R_0^{\pi(2)} \label{eq:holds_by_initial_assumption_2}
\end{IEEEeqnarray} 
where \eqref{eq:holds_by_initial_assumption_2} holds because  $\sigma_{\pi(1)}R_0^{\pi(1)}\geq 0$, because $I(U_0;Y_0)\geq R_0^{\pi(2)}$ by \eqref{eq:initial_assumption_R},  and $\sigma_0 + \sigma_{\pi(2)} \geq 1- \epsilon_{\pi(2)}$ by  \eqref{eq:sigmas}.

For $\sigma_{0} \leq  1-\epsilon_{\pi(1)}$, rate constraint \eqref{eq:R0une_simp2} can be verified as follows:
\begin{IEEEeqnarray}{rCl}
	R_0 &\geq& \sigma_{0}I(U_0;Y_0) + \sigma_{\pi(1)}R_0^{\pi(1)} + \sigma_{\pi(2)}R_0^{\pi(2)} \\
	&\geq & \sigma_{0}I(U_0;Y_0) + (1-\epsilon_{\pi(1)} - \sigma_{0})R_0^{\pi(1)} \nonumber \\
	&& +  \sigma_{\pi(2)}R_0^{\pi(2)} \label{eq:dd}\\
	&\geq& \sigma_{0}I(U_0;Y_0) + (1-\epsilon_{\pi(1)} - \sigma_{0})\big(R_0^{\pi(1)} + R_0^{\pi(2)}\big) \nonumber \\ 
	&& + (\epsilon_{\pi(1)}-\epsilon_{\pi(2)})R_0^{\pi(2)} \label{eq:holds_by_sigmas_2} \\
	&\geq& (1-\epsilon_{\pi(1)})I(U_0;Y_0) + (\epsilon_{\pi(1)}-\epsilon_{\pi(2)})R_0^{\pi(2)} \label{eq:holds_by_case3}
\end{IEEEeqnarray} 
where \eqref{eq:dd} holds by \eqref{eq:sigmas}, 
\eqref{eq:holds_by_sigmas_2} holds because $\sigma_{\pi(2)} \geq 1- \epsilon_{\pi(2)}- \sigma_{0}$ by \eqref{eq:sigmas}, and \eqref{eq:holds_by_case3} holds by \eqref{eq:condU} and $\sigma_0 \leq 1-\epsilon_{\pi(1)}$. This establishes that \eqref{eq:E3_simp1} holds under condition \eqref{eq:condU}.

The proof is concluded  by showing that for any tuple $(\theta_1,\theta_2, P_{U_0|Y_0}, R_0^1, R_0^2)$ satisfying \eqref{eq:E3_simp1}, \eqref{eq:initial_assumption}, and 
\begin{equation}
I(U_0;Y_0) > R_0^{\pi(1)} + R_0^{\pi(2)},
\end{equation} we can find a pmf $P_{\tilde{U}_0|Y_0},$ satisfying  \eqref{eq:E3_simp2} when $U_0$ is replaced by $\tilde{U}_0$.  
Choose a bivariate $\tilde{U}_0 = (\tilde U_0^1,\tilde U_0^2)$ such that  $\tilde U_0^1 \rightarrow Y_0 \rightarrow \tilde U_0^2$ forms a Markov chain and for each $i\in\{1,2\}$ the new random-variable $\tilde U_0^i$ achieves $\eta_{i}\big( R_0^{i}\big)$, i.e.,
\begin{equation}
R_0^{i} = I\big(Y_0;\tilde U_0^i\big) \quad \textnormal{and}\quad \eta_{i}\big(R_0^{i}\big) = I\big(\tilde U_0^i;Y_{i}\big). \label{eq:rate_eta_equality}
\end{equation}
Since for any $i\in\{1,2\}$ we have $I(\tilde{U}_0;Y_{i}) \geq I(\tilde{U}_0^i;Y_{i}) = \eta_{i}(R_0^{i})$, the exponents satisfy
\begin{IEEEeqnarray}{rCl}
	\theta_{\pi(1)} &\leq& \min \left \{I(U_0;Y_{\pi(1)}),\eta_{\pi(1)}\big(R_0^{\pi(1)}\big)\right \} = \eta_{\pi(1)}\big(R_0^{\pi(1)}\big)  \label{eq:s1}
	\IEEEeqnarraynumspace\\
	&\leq& {I(\tilde{U}_0;Y_{\pi(1)})}, \\
	\theta_{\pi(2)} &\leq& \min \big\{I(U_0;Y_{\pi(2)}),\eta_{\pi(2)}\big(R_0^{\pi(2)}\big)\big\} = \eta_{\pi(2)}\big(R_0^{\pi(2)}\big) \label{eq:s2} \\
	&=& \min\big\{I(\tilde{U}_0;Y_{\pi(2)}), \eta_{\pi(2)}\big(R_0^{\pi(2)}\big)\big\},
\end{IEEEeqnarray}
where the inequalities in \eqref{eq:s1} and \eqref{eq:s2} hold by \eqref{eq:initial_assumption}. Similarly,
\begin{IEEEeqnarray}{rCl}
	R_0 &\geq&  \sigma_{0}I(U_0;Y_0) + \sigma_{\pi(1)}R_0^{\pi(1)} + \sigma_{\pi(2)}R_0^{\pi(2)}\\
	&>& (1-\epsilon_{\pi(1)})R_0^{\pi(1)} + (1-\epsilon_{\pi(2)}) R_0^{\pi(2)}\label{eq:holds_by_condition0}\\
	& = &  (1-\epsilon_{\pi(1)})I\big(\tilde U_0^{\pi(1)};Y_0\big) + (1-\epsilon_{\pi(2)})I\big(\tilde U_0^{\pi(2)};Y_0\big) \label{eq:holds_by_the_markov_chainf}\\
	&\geq& (1-\epsilon_{\pi(1)})I\big(\tilde U_0^{\pi(1)};Y_0\big) + (1-\epsilon_{\pi(1)})I\big(\tilde U_0^{\pi(2)};Y_0|\tilde U_0^{\pi(1)}\big) \nonumber \\
	&& + (\epsilon_{\pi(1)}-\epsilon_{\pi(2)})I\big(\tilde U_0^{\pi(2)};Y_0\big) \label{eq:holds_by_the_markov_chain}\\
	&=& (1-\epsilon_{\pi(1)})I\big(\tilde{U}_0;Y_0\big) + (\epsilon_{\pi(1)}-\epsilon_{\pi(2)})I\big(\tilde U_0^{\pi(2)};Y_0\big) \IEEEeqnarraynumspace \label{eq:holds_by_definition_of_tilde_U0}
\end{IEEEeqnarray}
where inequality \eqref{eq:holds_by_condition0} holds by the assumption that $I(U_0; Y_0) > R_0^1 +R_0^2$ and by condition \eqref{eq:sigmas};  equality \eqref{eq:holds_by_the_markov_chainf}  holds by \eqref{eq:rate_eta_equality};  inequality \eqref{eq:holds_by_the_markov_chain} holds by the Markov chain $\tilde U_0^1 \rightarrow Y_0 \rightarrow \tilde U_0^2$; and \eqref{eq:holds_by_definition_of_tilde_U0} by the chain rule and the definition of $\tilde{U}_0$.\smallskip

\section{Strong Converse Proof to Theorem~\ref{thm2}}\label{strong_converse}

Fix an exponent pair in $\mathcal{E}_{\text{fix}}^*(R_0,R_1,R_2,\epsilon_1, \epsilon_2)$ and a sequence (in $n$) of encoding and decision functions $\{(\phi^{(n)}, g_1^{(n)}, g_2^{(n)})\}$ satisfying the constraints on the rate and the error probabilities in \eqref{type1constraint1}, \eqref{thetaconstraint}, \eqref{eq:maxRatei}.

Fix a positive $\eta>0$ and a blocklength $n$ and
choose  $\mu_n=n^{-2/3}$. Define for $i\in\{1,2\}$, the sets 
\begin{IEEEeqnarray}{rCl}\label{Bn_strongconverse}
	\mathcal{B}_{i}(\eta) &\triangleq& \{y_0^n \in \mathcal{T}_{\mu_n}^{(n)}(P_{Y_0}) \colon \nonumber \\
	&& \; \mathrm{Pr}[\hat{\mathcal{H}}_{Y_i}=0 | Y_0^n = y_0^n, \mathcal{H}=0] \geq \eta\}, \; i\in\{1,2\},\IEEEeqnarraynumspace
	\label{eq:Bnidef_strongconverse}\\
	\mathcal{D}_{0}(\eta) &\triangleq& 	\mathcal{B}_{1}(\eta) \cap 	\mathcal{B}_{2}(\eta).
\end{IEEEeqnarray}
Further define the probability
\begin{IEEEeqnarray}{rCl}
	\Delta_0 &\triangleq& P_{Y_0^n}(\mathcal{D}_{0}(\eta)),
\end{IEEEeqnarray}
and notice that by the laws of probability
\begin{IEEEeqnarray}{rCl}\label{eq:delta0_ub}
	\Delta_0 &\geq& P_{Y_0^n}(\mathcal{B}_{1}(\eta)) + P_{Y_0^n}(\mathcal{B}_{2}(\eta)) - 1.
\end{IEEEeqnarray}
By   \eqref{type1constraint1}, it can further be shown that
	\begin{IEEEeqnarray}{rCl}
		1 - \epsilon_i  &\leq&
		\eta(1- P_{Y_0^n}(\mathcal{B}_i(\eta))) + P_{Y_0^n}(\mathcal{B}_i(\eta)) +  P_{Y_0}^{n}\left(\overline{\mathcal{T}}_{\mu_n}^{(n)}\right).\IEEEeqnarraynumspace \label{eq:Bi_derivation_strong_converse} 
	\end{IEEEeqnarray}
	Thus, by \eqref{eq:Bi_derivation_strong_converse} and  \cite[Lemma~2.12]{Csiszarbook}:
\begin{IEEEeqnarray}{rCl}\label{eq:PB_strong_converse}
\Delta_0 &\geq& {1 - \epsilon_1 - \epsilon_2 - \eta \over{1 - \eta}} - {\vert{\mathcal{Y}_0}\vert \over{(1-\eta) \mu_n n}}, \IEEEeqnarraynumspace
\end{IEEEeqnarray}
and we  conclude that in the limit $n\to \infty$ and $\eta \downarrow 0$:
\begin{subequations} \label{eq:Delta_conditions_strong_converse}
	\begin{IEEEeqnarray}{rCl}
		\lim_{\eta \downarrow 0} \lim_{n\to \infty}\Delta_0 & \geq &  1- \epsilon_1 -\epsilon_2.
	\end{IEEEeqnarray}
\end{subequations}

We proceed by applying Lemma~\ref{lem:receiverconverse} to the set $\mathcal{D}_0$. By the initial condition $\epsilon_1 + \epsilon_2 <1$, and thus for $\eta>0$ sufficiently small and $n$ sufficiently large, by \eqref{eq:Delta_conditions_strong_converse} $\Delta_0$ is positive and  we can  apply Lemma~\ref{lem:receiverconverse} to the set $\mathcal{D}_0(\eta)$. By this Lemma~\ref{lem:receiverconverse}, and using also the maximum-rate constraints  \eqref{eq:maxRatei},  and the trivial inequality $nR_i \geq H(\tilde{M}_i)$, for all $i \in \{0,1,2\}$, we conclude that there is a tuple $(U_0,U_1,U_2)$ satisfying  
\begin{IEEEeqnarray}{rCl}
	nR_0 &\geq& H(\tilde{M}_{0}) \geq nI(U_0;\tilde{Y}_0) + \log P_{Y_0^n}(\mathcal{D}_0), \label{eq:M1i_stron_converse} \IEEEeqnarraynumspace\\
	nR_i &\geq& H(\tilde{M}_{i}) \geq nI(U_{i};\tilde{Y}_{0}|U_0), \quad i \in \{1,2\}, \IEEEeqnarraynumspace\label{eq:M2i_strong_converse}
\end{IEEEeqnarray}
and for $i \in \{1,2\}$ :
\begin{IEEEeqnarray}{rCl}
	-\frac{1}{n} \log \beta_{i,n}
	&& \leq I(U_0U_i;\tilde{Y}_i) + \o_{i}(n),\IEEEeqnarraynumspace \label{eq:sumI_strong_converse}
\end{IEEEeqnarray}
where for each $i$,  the function $\o_{i}(n)\to 0$ as $n\to \infty$ and the random variables $\tilde{Y}_{0},\tilde{Y}_i, \tilde{M}_0, \tilde{M}_{i}$ are defined as in the lemma applied to the set $\mathcal{D}_0$.


Thus we have proved so far that for all $n\geq 1$ there exists joint pmf $P_{U_0U_1U_2\tilde{Y}_0\tilde{Y}_1\tilde{Y}_2} = P_{\tilde{Y}_0}P_{\tilde{Y}_1\tilde{Y}_2|\tilde{Y}_0}P_{U_0U_1U_2|\tilde{Y}_0}$ (abbreviated as $P^{(n)}$) so that the following conditions hold for $i\in\{1,2\}$ 
\begin{subequations}\label{eq:conditions_strong_converse}
	\begin{IEEEeqnarray}{rCl}			
		R_0 &\geq  &  I_{P^{(n)}}({U}_0;\tilde{Y}_0) + g_{1}(n) ,\label{eq:R111_strong_converse} \IEEEeqnarraynumspace \\
		R_i &\geq  &I_{P^{(n)}}({U}_i;\tilde{Y}_0|U_0), \label{eq:R222_strong_converse}\\
		\theta_i &\leq& I_{P^{(n)}}({U}_0U_i;\tilde{Y}_i), + g_{2,i}(n), \label{eq:theta111_strong_converse}\IEEEeqnarraynumspace
	\end{IEEEeqnarray}
\end{subequations}
for some nonnegative functions $g_{1}(n), g_{2,i}(n)$ with the following asymptotic behaviors:
\begin{IEEEeqnarray}{rCl}
	\lim_{n\to \infty} g_{1}(n) &=& 0, \IEEEeqnarraynumspace\\ 
	\lim_{n\to \infty} g_{2,i}(n) &=& 0, \qquad  \forall i \in \{1,2\}.
\end{IEEEeqnarray}

The rest of the proof follows the same steps as the proof of the converse in Section~\ref{general_converse}.
By Carath\'eodory's theorem \cite[Appendix C]{ElGamal}, there exist for each $n$ random variables ${U}_0,U_1,U_2$ satisfying \eqref{eq:conditions_strong_converse} over alphabets of sizes
\begin{align}
	\vert {\mathcal{U}}_0\vert &\leq \vert \mathcal{Y}_0\vert + 3, \\
	\vert {\mathcal{U}}_i \vert &\leq \vert {\mathcal{U}}_0 \vert\cdot|\mathcal{Y}_0| + 1, \quad i \in \{1,2\}.
\end{align}
Invoke the Bolzano-Weierstrass theorem and  consider a sub-sequence  $P_{U_0U_1U_2\tilde{Y}_0\tilde{Y}_1\tilde{Y}_2}^{(n_k)}$ that converges to a limiting pmf $P_{U_0U_1U_2Y_0Y_1Y_2}^{*}$. For these limiting pmfs, which we abbreviate by $P^*$, we conclude by \eqref{eq:R111_strong_converse}--\eqref{eq:theta111_strong_converse} that for all $i\in\{1,2\}$:
\begin{IEEEeqnarray}{rCl}			
	R_0& \geq &  I_{P^{*}}({U}_0;{Y}_0) ,\label{eq:R_1_f_strong_converse} \\
	R_i &\geq  & I_{P^{*}}({U}_i;{Y}_0|U_0),\\
	\theta_i &\leq & I_{P^{*}}(U_0{U}_i;{Y}_i). \label{theta_1_f_strong_converse}
\end{IEEEeqnarray}
Notice further that since for any $k$, the sequence $\tilde{Y}_0^{n_k}$ lies in the typical set $\mathcal{T}^{(n_k)}_{\mu_{n_k}}(P_{Y_0})$, we have $\vert P^{(n_k)}_{\tilde{Y}_0} - P_{Y_0}\vert \leq \mu_{n_k}$ and thus the limiting pmfs satisfy $P^*_{Y_0}=P_{Y_0}$. Moreover, since for each $n_k$ the pair of random  variables $\big(\tilde{Y}_1,\tilde{Y}_2\big)$ is drawn according to $P_{Y_1Y_2|Y_0}$ given $\tilde{Y}_0$, the limiting pmf also satisfies $P_{Y_1Y_2|Y_0}^*=P_{Y_1Y_2|Y_0}$. 
We also notice that  under $P^*$ the  Markov chain
$(U_0,U_1,U_2)\to Y_0 \to (Y_1,Y_2)$ holds.  This concludes the proof.

\newpage
\bibliographystyle{ieeetr}
\bibliography{references}

\end{document}